\documentclass[12pt,preprint]{aastex}
\usepackage{natbib}
\usepackage{graphicx,graphics}
\usepackage{amsmath}

\newcommand{\HI}{\hbox{H\,{\sc i}}}

\newcommand{\raJ}[4]{$\alpha_{\mathrm{\scriptscriptstyle J2000.0}}~=~
  #1^{\mathrm{h}}#2^{\mathrm{m}}#3^{\mathrm{s}}\!\!.#4$}
\newcommand{\decJ}[4]{$\delta_{\mathrm{\scriptscriptstyle J2000.0}}~=~
  #1^{\circ}#2^{\prime}#3^{\prime\prime}\!\!.#4$}

\begin{document}
\title{Localized SiO emission triggered by the passage of the W51C SNR shock}

\author{G. Dumas$^{1}$, S. Vaupr\'e$^2$, C. Ceccarelli$^2$,
  P. Hily-Blant$^2$, G. Dubus$^2$, T. Montmerle$^3$, S. Gabici$^4$}
\altaffiltext{1}{IRAM, Grenoble, F-38041, France}
\altaffiltext{2}{UJF-Grenoble 1 / CNRS-INSU, Institut de Plan\'{e}tologie et 
d\textquoteright Astrophysique de Grenoble (IPAG) UMR 5274, Grenoble, F-38041, France}
\altaffiltext{3}{Institute d'Astrophysique de Paris, France}
\altaffiltext{4}{ APC, AstroParticule et Cosmologie, Universit\'e Paris Diderot, CNRS, CEA, Observatoire de
Paris, Sorbonne Paris , France}
    \date{Received - ; accepted -}
\begin{abstract}

The region towards W51C is a convincing example of interaction between a supernova remnant and a surrounding molecular cloud. Large electron abundances have been reported towards the position W51C-E located in this interaction region, and it was proposed that the enhanced ionization fraction was due to cosmic ray particles freshly accelerated by the SNR shock. We present PdB interferometer observations of the H$^{13}$CO$^+$(1-0) and DCO$^+$(2-1) emission lines centered at position W51C-E. These observations confirm the previous scenario of cosmic-ray induced ionization at this location. In addition, SiO(2-1) emission has been successfully mapped in the close vicinity of W51C-E, with a spatial resolution of 7\arcsec. The morphology and kinematics of the SiO emission are analyzed and strongly suggest that this emission is produced by the passage of the SNR primary shock. Put in conjunction with the enhanced ionization fraction in this region, we give a consistent picture in which the W51C-E position is located downstream of the shock, where a large reservoir of freshly accelerated particles is available.
\end{abstract}

\section{Introduction}

Cosmic Rays (CR) permeate our Galaxy, playing a major role in setting the prevalent physical conditions, notably because they are responsible for the ionization of the dense part of the molecular clouds where stars form. Galactic CRs are thought to originate mostly from the shock of supernovae remnants (SNR) where they can be accelerated up to PeV energies. Their presence can be inferred from the effects of their interaction with their environment.  High energy
CRs ($\gtrsim 1$~GeV) can be tracked down because, when they hit a large mass of gas, they
cause bright $\gamma$-ray emission, due to inelastic proton--proton interactions (Kelner et al. 2006). On the other hand, low energy ($\lesssim 1$~GeV) CR can be probed by the enhancement
of the ionization that they cause in nearby molecular gas (Padovani et al. 2009). 
In fact, ionization is the only way to probe the low-energy end of the CR spectrum. Hence, SNR interacting with molecular clouds are particularly promising sites to study both the properties of freshly accelerated CR (e.g. density, spectrum) and their feedback on their environment (diffusion, ionization). In this context, the presence of both an enhanced ionization degree and gamma-ray emission associated with dense molecular clouds close to SNR  can provide additional evidence of CR acceleration from sub-GeV (ionization) to multi TeV (VHE gamma-rays) energies. 

\begin{figure}
\begin{center}
\includegraphics[height=7cm]{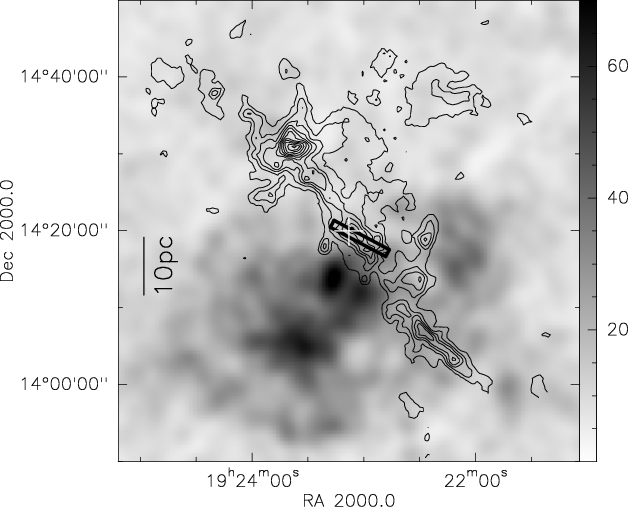} 
\includegraphics[height=7cm]{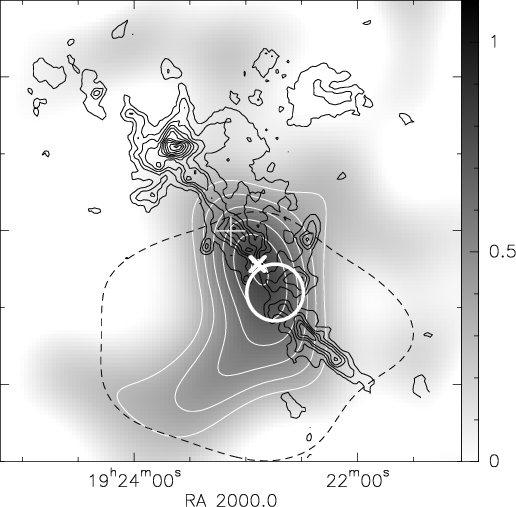} 
\caption{The high energy environment of W51C. In both panels, the large white cross locates the phase center of our PdBI observations at the position of W51C-E and the $^{13}$CO(2-1) integrated intensity between 60 and 75\,km\,s$^{-1}$ is shown as black contours (from 5 to 60\,K\,km\,s$^{-1}$ in s.jpeg of 5\,K\,km\,s$^{-1}$; Jackson et al. 2006).  \emph{left}: ROSAT X-ray map of SNR W51C (in counts s$^{-1}$, Koo et al. 1995) is shown in gray scale. The black rectangle delineates the field-of-view of our IRAM 30m observations. \emph{right}: {\it MAGIC} gamma-ray emission map from 300~Gev to 1~TeV (in relative flux, Aleksi\'{c} et al. 2012), shown in gray scale. The white plain contours  give the {\it MAGIC} relative flux from 0.3 to 0.8 in s.jpeg of 0.1 and the dashed contour marks the 260 counts deg$^{-2}$ measured by the {\it Fermi}/LAT in the 2-10 GeV energy range. The white cross corresponds to the position of the OH maser emission (Koo et al. 1995, Green et al. 1997) and the white circle shows the region of shocked atomic and molecular gas (Koo \& Moon 1997a,b).  }
\label{fig:global}
\end{center}
\end{figure}

The first dense molecular cloud where the association of enhanced ionization and gamma-ray emission has been detected is associated with the SNR W51C (Ceccarelli et
al. 2011 hereinafter CHMD2011). W51C is  $\sim$5\degr\ away from the galactic plane, at a distance of $\sim$5.5\,kpc (Cyganowski et al. 2011). 
It extends about 30\arcmin\ and its age is $\sim3\times10^4$ yrs
(Koo et al. 1995). Figure~\ref{fig:global} presents the environment of W51C. The thermal X-ray emission (Fig.~\ref{fig:global} left panel), tracing the SNR primary shock, extends north-west beyond the molecular cloud, hinting that the SNR is situated behind the W51B cloud complex (Koo et al. 1995).  A bright and extended GeV-TeV $\gamma$-ray source has been detected towards W51C by Feinstein et al. (2009), Abdo et al.  (2009) and Aleksi\'{c} et al. (2012) who concluded that the origin of such emission is the interaction between the SNR and the nearby molecular cloud (Fig.~\ref{fig:global}, right panel). OH masers and high velocity \HI\ and molecular emission are observed coincidently in the overlapping region of W51C and the W51B complex, which suggests that they are associated with the SNR shock  (Koo et al. 1995, Green et al. 1997, Brogan et al. 2000), although that is still under debate (Tian \& Leahy 2013).

CHMD2011 sampled several lines-of-sight toward 
this interaction region  with the IRAM-30m single dish telescope. They found
that the CR ionization rate is about 100 times larger than in 
standard galactic molecular clouds in the direction of the position
W51C-E  \raJ{19}{23}{08}{0}; \decJ{14}{20}{00}{0}, situated north west of the SNR, at the edge of the gamma-ray emission detected by {\it Fermi}/LAT and {\it MAGIC} as shown in Fig.~\ref{fig:global} (right panel) and about 2\arcmin\ north of the H{\sc ii} region G49.2-0.3. The CR ionization rate was derived by measuring the abundance ratio
DCO$^+$/HCO$^+$, following the method described in Gu\'elin et al. (1977, see CHMD2011 for further details on the  method used).  The
observations were sensitive to the ionization of the gas in the 28\arcsec\
beam, corresponding to a linear distance of $\sim 1$pc.

 Promptly after
the publication of this measurement, NIR observations of the region
revealed the presence of a protostar at the edge of the beam of the
DCO$^+$ and HCO$^+$ observations (Cyganowski et al. 2011), which could
alter the interpretation of the results of the CHMD2011 work. In practice, CHMD2011 assumed that all the detected DCO$^+$ and HCO$^+$ emission come from ionization of the dense gas by the CR. However, UV flux from a star formation region and potential outflows could also ionize the surrounding molecular gas. Therefore, if CHMD2011 single-dish observations were contaminated by emission towards the protostar, the beam-averaged measured ionization rate may not be related to the CR. Therefore, we carried
out high spatial resolution observations of the same region with the
Plateau de Bure Interferometer (PdBI) in order to verify whether the spatial distribution of 
DCO$^+$ and HCO$^+$ emission, used to derive the high CR ionization rate,  is linked to the protostar.

We describe here the results of these observations that provide additional evidence for an overdensity of freshly accelerated low energy CR in W51C-E. In addition we discuss the shocked regions probed by the SiO(2-1) emission detected by these PdBI observations.

\section{Observations and data reduction}

W51C-E was observed with Plateau de Bure Interferometer between June
and July 2012 in the D configuration with 5 antennae, providing
baselines between 15\,m and 111\,m. We used the 2mm receivers tuned at 144.077GHz and the 3mm receivers
tuned at 86.754GHz, to map the emission lines DCO$^{+}$(2-1) and
H$^{13}$CO$^{+}$(1-0), respectively. A total bandwidth of about
230\,MHz with a spectral resolution of 0.156\,MHz was used for both
set-ups. At 3mm, the set-up covers the SiO(2-1) line at 86.847\,GHz.
In parallel, the wideband correlator WideX offers a total of 3.6\,GHz
of bandwidth at 1.95\,MHz resolution. The phase center of the
observations was at \raJ{19}{23}{08}{0}; \decJ{14}{20}{00}{0}. MWC349 was used
as flux calibrator and the quasar J1923+210 was used as phase
calibrator and observed every 23min. At 3\,mm (respectively 2\,mm), the primary beam is 58\arcsec\  (respectively 32\arcsec) with a conversion factor of 22\,Jy\,K$^{-1}$  (respectively 29\,Jy\,K$^{-1}$). 

The data reduction was done with the standard IRAM GILDAS\footnote{http://www.iram.fr/IRAMFR/GILDAS} software
packages CLIC and MAPPING (Guilloteau \& Lucas 2000). Each day of
observations was calibrated separately and single datasets were
created at 2\,mm and 3\,mm for both the narrow band correlator and
WideX. Pure continuum and continuum subtracted data cubes were then
created and deconvolved separately, using natural weighting leading to
a beam size of $3\farcs8 \times 3\farcs4$ at 2\,mm and $7\farcs4
\times 4\farcs4$ at 3\,mm.  In the case of the WideX data, all
detected emission lines were cleaned independently. The final natural weighted data cubes  have a rms of 4.8\,mJy\,bean$^{-1}$  in a channel width of 1.3\,km\,s$^{-1}$ (0.625\,MHz) at 2\,mm and 1.1\,km\,s$^{-1}$ (0.3125\,MHz) at 3\,mm.

%
%
%
%
%
%

In addition, we obtained a $10'$x$2'$ map of the SiO(2-1) line with the
IRAM-30m telescope.  The observations were carried out in January 2013.  We used simultaneously the EMIR bands E090 and E230 and tuned
the Fast Fourier Transform Spectrometer (FFTS) backends on H$^{13}$CO$^{+}(1-0)$ and $^{13}$CO$(2-1)$ transitions, respectively.
The mapping of the region was conducted in the on-the-fly mode over $\sim 4$ hours, reaching a rms of 90\,mK in 0.67\,km\,s$^{-1}$ channel width at 3\,mm and 275\,mK  in 0.27\,km\,s$^{-1}$ channel width  at 1\,mm. The weather was good and $T_\text{sys}$ values were typically lower
than 150\,K at 3\,mm and 350\,K at 1\,mm.

We used a reference position about  $\delta$RA=300\arcsec, $\delta$dec=-400\arcsec\ away  from the mapped area.
The amplitude calibration was done typically every 15 minutes,
and pointing and focus were checked every 1 and 3 hours,
respectively, ensuring $\approx$2\arcsec\ pointing accuracy. All spectra were
reduced using the CLASS package (Hily-Blant et al. 2005) of the
IRAM GILDAS software. 
Residual bandpass effects were subtracted using low-order ($\leq 3$)
polynomials. Table~\ref{tab:obs} lists the set-ups used and the emission lines mapped with these single-dish and interferometric observations.

\section{Results}

\subsection{DCO$^+$ and H$^{13}$CO$^+$}

We find no compact DCO$^+$ or H$^{13}$CO$+$ source within the beam covered by previous IRAM-30m observations of W51C-E  by CHMD2011.  Therefore no emission associated to the nearby protostar is contaminating the measurements by CHMD2011. This result supports the
conclusions by these authors that the entire region traced by the 30m
beam has an enhanced CR ionization flux.

\subsection{SiO(2-1)}

\begin{figure}
\begin{center}
\includegraphics[height=5cm]{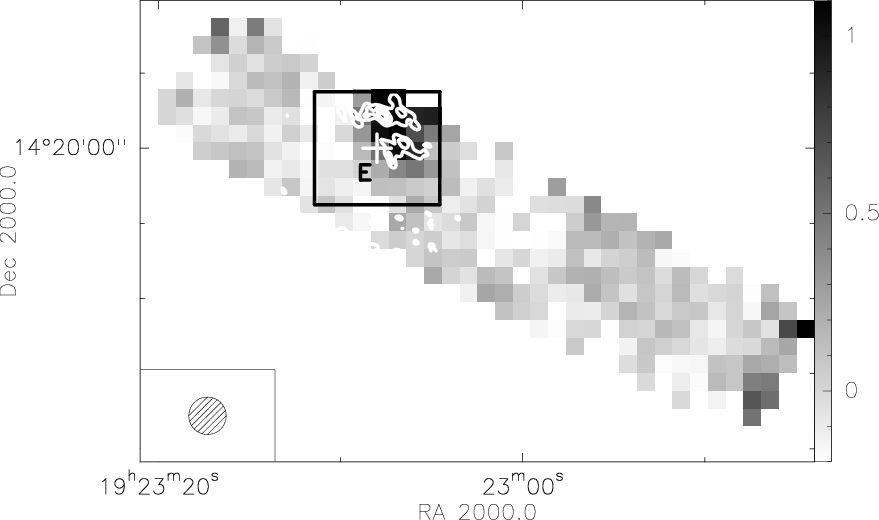} 
\includegraphics[height=5cm]{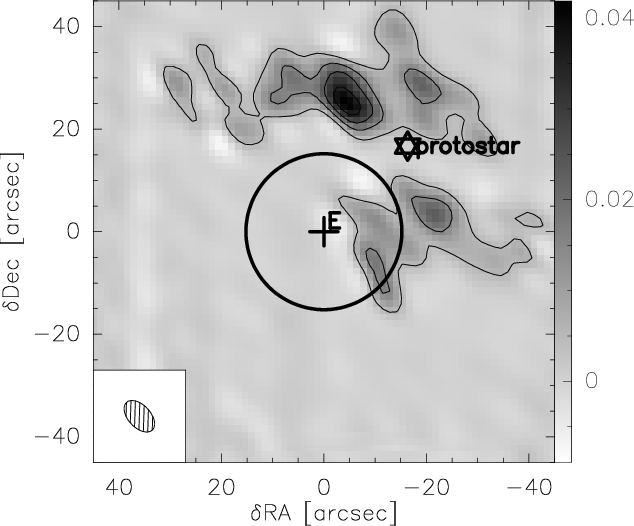} 
\caption{\emph{left}: SiO(2-1) map from the IRAM 30m telescope (K\,km\,s$^{-1}$, in antenna temperature scale T$_A^*$. The rms is 60\,mK\,km\,s$^{-1}$) overlaid with the SiO(2-1) emission observed with the PdBI (white contours from 0.005 to 0.05 in s.jpeg of 0.01 Jy\,beam$^{-1}$\,km\,s$^{-1}$).  The white cross indicates the W51C-E position and the black square represents the size of the right panel.  The left bottom panel shows the 30m half power beam of $28\arcsec$. \emph{right}: PdBI SiO(2-1) integrated emission (in Jy\,beam$^{-1}$\,km\,s$^{-1}$, rms=4.8\,mJy\,beam$^{-1}$). The contours are from 0.005 to 0.05 in s.jpeg of 0.01 Jy\,beam$^{-1}$\,km\,s$^{-1}$. The black circle represents the primary beam of the IRAM 30m telescope at 3\,mm (28\arcsec), centered on the W51C-E position (cross). The black star marks the position of the protostar observed in the IR and cm wavelengths (Cyganowski et al. 2011). The CLEAN beam of $7\farcs4 \times 4\farcs4$  is shown in the bottom left corner.}
\label{fig:SiO-intensity}
\end{center}
\end{figure}

\begin{figure}
\begin{center}
\includegraphics[width=10cm]{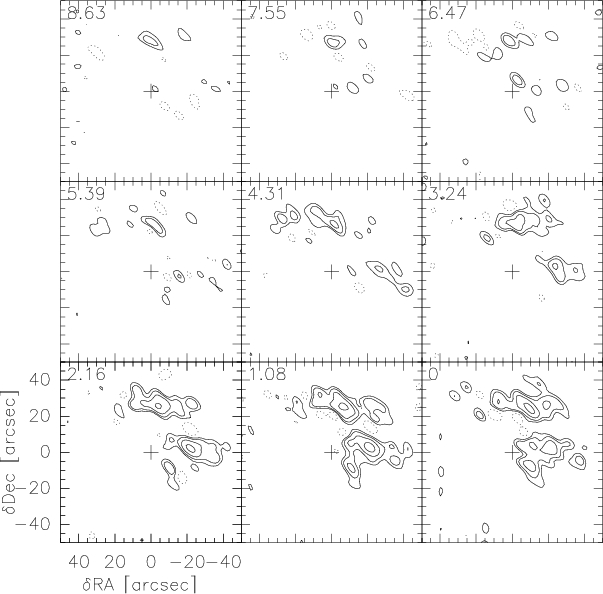} 
\caption{Channel maps of the naturally weighting SiO(2-1) emission towards W51C-E. The channels are 1.08 km s$^{-1}$ wide between $\sim -8.63$ and $\sim +8.63$ km s$^{-1}$ around the LSR velocity ($v_{lsr}=67$ km s$^{-1}$). Contours are plotted at  $-3\sigma$, $3\sigma$, $6\sigma$, $12\sigma$ and $24\sigma$ with 1$\sigma = 4.8$ mJy beam$^{-1}$. The CLEAN beam of $7\farcs4 \times 4\farcs4$  is shown in the bottom left corner and the cross marks the phase center of the observations.}
\label{fig:SiO-channels}
\end{center}
\end{figure}

\begin{figure}
\begin{center}
\includegraphics[width=10cm]{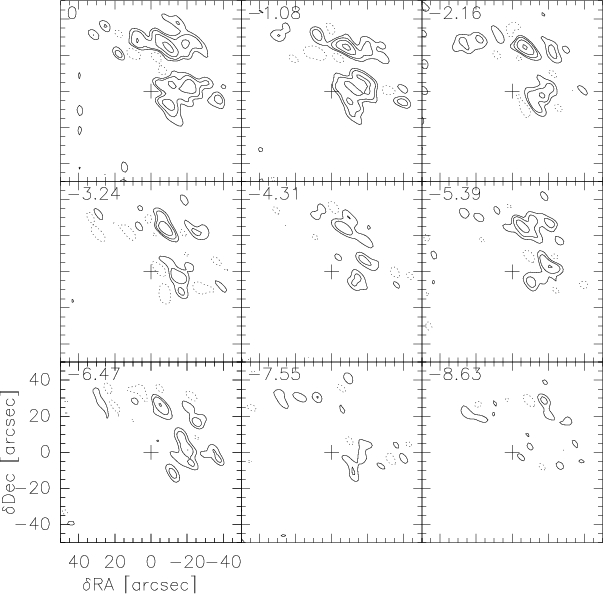} 
\caption{Figure.~\ref{fig:SiO-channels} continued}
\label{fig:SiO-channels2}
\end{center}
\end{figure}

Figure \ref{fig:SiO-intensity}, left panel shows the IRAM-30m map of the SiO
emission. Strong emission is detected only around the position of the
W51C-E source.
The PdB SiO integrated map around that position is shown in the right panel of Fig.~\ref{fig:SiO-intensity}.
The SiO emission is concentrated in two regions, north and south of
the protostar, respectively. The north region is split into three
clumps, aligned in the west-east direction over a length of about 1.2\,pc
($\sim$45\arcsec). The southern SiO structure is weaker and splits into
two emitting regions with about a 45 deg angle and an elongation of about
0.5\,pc ($\sim$20\arcsec). These SiO clumps are barely resolved, with an average size of about 0.3\,pc (10\arcsec),  and have a velocity dispersion between 5 and 7\,km\,s$^{-1}$.  
Figures~\ref{fig:SiO-channels} and \ref{fig:SiO-channels2} present the channel maps of the SiO
emission, at 1.08 km s$^{-1}$ velocity resolution. The first thing to
note, no velocity gradient nor structure apparently associated with
the protostar is evident in these maps. The emission is only present from -6.5 to +6.5\,km\,$^{-1}$ around the systemic velocity (67\,km\,s$^{-1}$, CHMD2011) with almost constant
intensity in both north and south structures.
%

\section{Discussion}

The first important result of this work is that the position observed
by CHMD2011 does not have any compact region of
DCO$^+$ emission, which supports the idea that the whole region
encompassed by the IRAM-30m beam is permeated by gas with a CR ionization
rate about 100 times larger than the standard one. In other words, the
protostar in the field has no impact on the determination of the ionization of the region.

The second, unexpected result is the presence of SiO emission,
concentrated in that region. As shown from the large scale map (Fig.~\ref{fig:SiO-intensity}, left panel), SiO is
only associated with W51C-E, which adds support to the conclusions by
CHMD2011 that it is a  peculiar region.
The obvious questions are: why is SiO present, what does it trace, and
has it anything to do with the enhanced CR ionization rate in the region ?
Under typical dark cloud conditions, silicon is almost tied up in interstellar grains, and its abundance in the gas phase is typically $N({\rm SiO})/N({\rm H_2}) \sim 10^{-11}$ or less (see Lucas \& Liszt 2000 and references therein){lucas2000}, because Si is trapped in the refractory
interstellar grains and/or mantles that envelop them. However,
in shocked regions, the grains can be shattered and the mantles be
sputtered (Flower et al. 1996, Schilke et al. 1997, Gusdorf et al. 2008),
 so that SiO, the major Si-bearing reservoir in molecular
gas, becomes orders of magnitude more abundant, up to $\sim
10^{-6}$. In fact, SiO is commonly used to trace molecular shocks
(Bachiller 1996).  
Our observations do not allow us to give an estimate of the abundance
nor of the physical conditions of the gas emitting the SiO(2-1) transition. However,
it is likely that the two observed structures have an enhanced SiO
abundance with respect to the rest of the whole region (over the
$10'$x$2'$ length of the IRAM-30m map). Even though we cannot quantify
the strength of the shock, there is no doubt that two shocked regions are present, probed by the SiO emission.
The next question is: what is the origin of this shock? One would
first think that it is powered by the protostar outflow. To check this possibility, we show the SiO emission channel maps in Fig.~\ref{fig:SiO-channels} and Fig.~\ref{fig:SiO-channels2}. The spatial shift between the red and blue shifted emission parts, which is a signpost of protostellar outflow (e.g. Gueth \& Guilloteau 1999) is not seen. 
Therefore, we conclude that the observed SiO emission is not
associated with an outflow driven by this star. This is in agreement with
previous studies detecting no outflow associated with the protostar (de Buizer \& Vacca 2010, Cyganowski et al. 2011)

Then the second more plausible explanation is that the detected shock
is connected, in a way or another, to the SNR primary shock that may have created
the large flux of CR probed by the enhanced ionization in W51C-E. 
Fig.~\ref{fig:global} shows the high energy environment of the SNR. In particular the W51C-E position (black cross in the right panel of Fig.~\ref{fig:global}) lies at the edge of the gamma-ray emission detected by {\it Fermi}/LAT and {\it MAGIC}. This leads us to conclude that the W51C-E position where enhanced ionization has been detected is situated in the vicinity of the SNR shock.
In general, SNR shocks are believed to be the main accelerators of Galactic CRs. Their velocities exceed $10 000$~km s$^{-1}$ in the initial phase of the evolution of SNRs, and gradually decrease as the SNRs age (Vink 2012). Koo et al. (1995) derived a forward shock velocity of $\approx 500\rm\,km\,s^{-1}$ and a dynamical age of $\approx3\times10^4$ years from the thermal X-ray emission. 
 Even at this late stage, shocks are believed to be strong enough to accelerate CRs up to GeV energies and possibly more (Ptuskin \&  Zirakashvili 2003).
 Theory predicts that if molecular clumps of dense gas obstruct the passage of the primary shock, its propagation stalls while a complex system of shocks develops in and around the clump (e.g. Jones \& Kang 1993, Inoue et al. 2012).  The localized SiO emission that we detect in W51C-E could thus arise from a cloud overrun by the SNR.  Assuming a primary shock speed $v_a\approx 500\rm\,km\,s^{-1}$, a cloud overdensity with respect to the ambient density  $n_c/n_a\la 40$ would give a typical transmitted shock speed in the clump $v_c \approx v_a (n_a/n_c)^{1/2}\ga 25\rm\,km\,s^{-1}$, enabling grain  sputtering (Gusdorf et al. 2008). If correct, this interpretation implies that  W51C-E is located downstream of the shock, where a large reservoir of ionizing CRs is available. This is further supported by the general picture that the primary shock, as traced by the thermal X-ray emission, extends north west beyond the W51B complex (Fig.~\ref{fig:global}). Only a part of the SNR is interacting with the north east portion of the W51B complex, as evinced from the presence of shocked atomic and molecular material coincident with the centroid of the gamma-ray emission, about 20\,pc (10\arcmin) south-west from our observations (Koo \& Moon 1997a, 1997b, Brogan et al. 2013).
 In principle, the shock traced by SiO emission might also be responsible for the enhanced CR ionization rate of W51C-E. Reflected shocks are expected to be significantly weaker than the primary shock, hence less efficient in accelerating CRs unless the primary shock is CR dominated (Jones \& Kang 1993). Though this possibility cannot be ruled out here, it seems less natural as the low-energy CR would also have to escape the clump rapidly to fill in the $\approx 1$ pc region where we detect an overionization. Probably,
a way to test this explanation is observing a gradient of CR
ionization rate in positions close to W51C-E: if the SiO shock is responsible for the CR acceleration, then the larger the distance
from the SiO shocks, the lower the ionization.

\section{Conclusions}

We have presented PdBI observations of the position W51C-E  in the interacting region of  the SNR W51C with the nearby molecular cloud. Our first result is that we confirm that this region studied by CHMD2011  is
  filled up with gas ionized by a enhanced flux of CRs. We also report the presence of two structures probed by SiO(2-1) emission,
  approximately elongated in the east-west direction, and we propose that these structures are tracing a low-velocity shock caused by the passage of the primary SNR shock through a dense clump. We conclude that the W51C-E region  is downstream of the SNR forward shock, exactly where the CRs accelerated at the forward shock would accumulate  and cause the enhanced ionization measured there. Although less probable, it is possible that the shock traced by the SiO emission
  is responsible for the acceleration of low ($\leq$~1\,GeV) CRs, which
  would contribute to the ionization of the dense gas in this region.

\begin{acknowledgements}
  This work has been supported by the french ``Programme National Hautes Energies''.  SG acknowledges the financial support of the UnivEarthS Labex program at Sorbonne Paris CitŽ (ANR-10-LABX-0023 and ANR-11-IDEX-0005-02). We wish to thank the staff at the IRAM 30m telescope and on the Plateau de Bure. We thank also the referee for his/her helpful comments. 

\end{acknowledgements}


%

%


\begin{table}
 \caption{observational parameters of the PdBI and IRAM-30m observations}
 \label{tab:obs}
\centering
\begin{tabular}{|l|c|c|}
\hline
\hline
 PdBI observations & 2\,mm & 3\,mm  \\ \hline
central frequency  & 144.077\,GHz &86.754\,GHz  \\ \hline
beam size& $3\farcs8 \times 3\farcs4$ &  $7\farcs4 \times 4\farcs4$ \\ \hline
velocity resolution& 1.3\,km\,s$^{-1}$ &  1.1\,km\,s$^{-1}$\\ \hline
rms (mJy beam$^{-1}$)& 3.1 &  4.8 \\ \hline
emission lines & DCO$^+$(2-1)& H$^{13}$CO$^+$(1-0); SiO(2-1)\\ \hline
primary beam &32\arcsec& 58\arcsec\\ \hline
conversion factor (Jy\,K$^{-1}$)& 29& 22\\
\hline
\hline
30m observations & 1\,mm & 3\,mm  \\ \hline
central frequency  & 220.399\,GHz &86.754\,GHz  \\ \hline
 beam  size&11\arcsec& 28\arcsec\\ \hline 
velocity resolution& 0.27\,km\,s$^{-1}$ &  0.67\,km\,s$^{-1}$\\ \hline
rms (mK)& 275 & 90 \\ \hline
emission lines & $^{13}$CO(2-1)& H$^{13}$CO$^+$(1-0); SiO(2-1)\\ \hline
\hline
\end{tabular}
\end{table}

\end{document}